\documentclass[12pt]{article}

\usepackage{amsmath, amsthm}
\usepackage{graphicx}
\usepackage{enumerate}
\usepackage{natbib}
\usepackage{url} 
\usepackage{color}
 \usepackage{multirow}

\usepackage{adjustbox} 
\usepackage{array}
\newcolumntype{L}{>{\centering\arraybackslash}m{3cm}}
\newcommand{\blind}{0}

\begin{document}

\def\spacingset#1{\renewcommand{\baselinestretch}%
{#1}\small\normalsize} \spacingset{1}


\if0\blind
{
  \title{\bf An Empirical Comparison of Parametric and Permutation Tests for Regression Analysis of Randomized Experiments}
\author{Kellie Ottoboni \\
Department of Statistics; Berkeley Institute for Data Science\\
University of California, Berkeley \\ [.1in]
Fraser Lewis \\
Medical Affairs and Evidence Generation \\
Reckitt Benckiser\\  [.1in]
Luigi Salmaso\\
Department of Management and Engineering \\
University of Padova
}  \maketitle
} \fi

\if1\blind
{
  \bigskip
  \bigskip
  \bigskip
  \begin{center}
    {\LARGE\bf  An Empirical Comparison of Parametric and Permutation Tests for Regression Analysis of Randomized Experiments}
\end{center}
  \medskip
} \fi

\bigskip
\begin{abstract}
Hypothesis tests based on linear models are widely accepted by organizations that regulate clinical trials.
These tests are derived using strong assumptions about the data-generating process so that the resulting inference can be based on parametric distributions.
Because these methods are well understood and robust, they are sometimes applied to data that depart from assumptions, such as ordinal integer scores.
Permutation tests are a nonparametric alternative that require minimal assumptions which are often guaranteed by the randomization that was conducted.
We compare analysis of covariance (ANCOVA), a special case of linear regression that incorporates stratification, to several permutation tests based on linear models that control for pretreatment covariates.
In simulations of randomized experiments using models which violate some of the parametric regression assumptions,
the permutation tests maintain power comparable to ANCOVA.
We illustrate the use of these permutation tests alongside ANCOVA using data from a clinical trial comparing the effectiveness of two treatments for gastroesophageal reflux disease.
Given the considerable costs and scientific importance of clinical trials, an additional nonparametric method, such as a linear model permutation test, may serve as a robustness check on the statistical inference for the main study endpoints.
\end{abstract}

\noindent%
\vfill

\newpage
\spacingset{1.45} 
\section{Background}
A hypothesis test is a statistical method for determining whether observed data is consistent with a belief about the process that generated the data.
Medical experiments use hypothesis testing to assess the evidence that a treatment affects one or more clinically relevant outcomes.
The simplest version of this experiment has been studied for nearly a century (see \cite{fisher_design_1935} and \cite[1990 translation]{neyman_application_1923} for early references).
This experiment involves randomly assigning two treatments to a fixed number of individuals in a group and measuring a single outcome.
One can conduct hypothesis tests and construct confidence intervals for the estimated treatment effect
by exploiting the fact that the difference in average outcomes between the two treatment groups is asymptotically normal with a variance that can be estimated from the data.

The goal of a randomized experiment is to draw causal inferences about the efficacy of a treatment.
Thus, it makes sense to analyze experiments in the context of potential outcomes, where every individual has a counterfactual outcome for each of the treatments (\cite{holland_1986_statistics})
Experiments are an attempt to estimate the counterfactual outcomes by randomly assigning treatments.
Random assignment of treatment ensures that pretreatment covariates are balanced between treatment groups on average, across all possible randomizations, making treatment groups comparable to each other.

However, in any particular randomization, there may be imbalances.
If the imbalanced variables are associated with the outcome, then even when treatment has no effect, there may be differences in outcomes between treatment groups.
Adjusting for such covariates can reduce the variability of treatment effect estimates and yield more powerful hypothesis tests.

Stratification, sometimes called blocking, is one method to control for covariates that are known a priori to be associated with the outcome.
Strata are groups of individuals with similar levels of a covariate.
These groups are defined during the design stage (i.e. before outcome data are collected).
Random assignment of treatments is conducted within each stratum, independently across strata.
This guarantees that the stratification variable is balanced between treatment groups.
A common stratification variable in clinical experiments is location:
individuals often come from many locations because it is difficult to recruit a sufficient number of participants at one site,
especially when the object of study is a rare disease or a rare outcome.

Linear regression is another method to control for imbalanced baseline covariates.
It is done during the data analysis stage.
In its simplest form, linear regression projects the outcomes onto the plane that best summarizes each variable's relationship to the outcome.
The coefficient of any particular covariate answers the question, if we were to hold fixed all other variables and increase this variable by one unit, how much would we expect the outcome to change?
This model posits a linear relationship between covariates, treatment, and outcome; 
if the true relationship is not linear, then linear regression gives the best linear approximation to the conditional expectation of the outcome.
It is standard to use analysis of covariance (ANCOVA) to incorporate stratification in a linear model.
ANCOVA is a particular case of linear regression that allows the mean outcome to vary from stratum to stratum.
This amounts to fitting a plane to each stratum, with the constraint that they have a common slope.

Hypothesis testing of estimated coefficients requires even stronger assumptions which are not guaranteed by the experimental design.
When the assumptions hold, a hypothesis test for a treatment effect amounts to a hypothesis test of the coefficient for treatment in the ANCOVA model, and can be evaluated analytically and efficiently.
A fully saturated linear model can yield asymptotically consistent estimates and confidence intervals (\cite{lin_agnostic_2013}).
This is especially problematic in medical trials, where conditions are not always ideal for the linear model to work well:
linear models can have substantial bias in small samples (\cite{freedman_regression_2008}),
outcomes are often discrete or ordinal, 
the treatment may have a differential effect across subgroups of individuals, 
and when the distributions of outcomes may differ across strata.

Permutation testing is an alternate approach (\cite{fisher_design_1935, pitman_significance_1937,pitman_significance_1938}).
Deliberate randomization induces a distribution for any test statistic under the null hypothesis that treatment has no effect on the outcome:
the randomization scheme provides information about all possible ways that treatment may have been assigned 
and the null hypothesis tells us what each individual's response would be regardless of the assignment (namely, it would be the same).
One determines how ``extreme'' the observed test statistic is relative to this randomization distribution, rather than a parametric reference distribution like Student's $t$ or the standard Gaussian.
Such a test is exact, meaning that it controls the type I error rate at the pre-specified level even in finite samples, whereas parametric hypothesis tests based on asymptotic approximations do not always guarantee good finite sample properties.
Permutation tests condition on the observed sample and do not require any assumptions about the way individuals were sampled from a larger population.
This is useful when the sampling frame is difficult to specify, such as when the study uses a convenience sample.

In the past, statisticians relied on parametric methods because asymptotic approximations were a computationally feasible way to estimate distributions and construct confidence intervals.
Now, computational power is no longer a barrier to finding exact (or exact to pre-specified precision) randomization distributions and confidence intervals.
In most cases, a randomization test is the ``gold standard'':
``[a] corresponding parametric test is valid only to the extent that it results in the same statistical decision [as the randomization test]'' (\cite{bradley_distribution_1968}).
There is no hard and fast rule describing the rate at which parametric tests approach the exact permutation solution, as they are both highly dependent on the particular data observed.
However, if the permutation test agrees with the parametric test, one may have a greater degree of confidence in the estimates and confidence intervals constructed using the parametric method.

We review several hypothesis tests for randomized experiments which adjust for pretreatment covariates to increase power to detect a nonzero treatment effect.  
We focus on ANCOVA and its permutation counterparts, comparing their performance in different scenarios and illustrating their application with a clinical dataset.
Section~\ref{sec:methods} introduces the potential outcomes model, shows how this model is inconsistent with the assumptions of the parametric ANCOVA, and describes permutation tests whose assumptions match this model.
Section~\ref{sec:simulations} presents simulations that suggest that even in this potential outcomes framework when various assumptions for the ANCOVA test are violated, the parametric and permutation tests have comparable power to detect a treatment effect.
In Section~\ref{sec:results}, we apply each of the tests to data from a clinical trial comparing the performance of two treatments for gastroesophageal reflux disease (GERD).
We conclude in Section~\ref{sec:discussion} with implications of these results for practitioners.

\section{Methods}\label{sec:methods}

\subsection{Notation}
Suppose we have a finite population of $N$ individuals.
Individuals are grouped into strata indexed by $j = 1, \dots, J$, with $n_j$ individuals in stratum $j$ and $\sum_{j=1}^J n_j = N$.
Of the $n_j$ subjects in stratum $j$, $n_j^T$ are assigned treatment 1, while the remaining $n_j - n_j^T$ are assigned treatment 0.
Let $Z_{ij}$ indicate the treatment assigned to individual $i$ in stratum $j$.

All individuals have two potential outcomes, $Y_{ij}(1)$ and $Y_{ij}(0)$, representing their responses to treatments $1$ and $0$, respectively.
We can never observe both; random assignment of treatment reveals $Y_{ij} = Z_{ij}Y_{ij}(1) + (1-Z_{ij})Y_{ij}(0)$.
The potential outcomes are fixed, but the observed outcome $Y_{ij}$ is random.
Throughout, we assume that there is no interference between individuals 
(in other words, $Y_{ij}$ is a function of $(Z_{ij}, Y_{ij}(1), Y_{ij}(0))$ and not any other $Z_{i', j'}$ for $(i', j') \neq (i, j)$)
and that there is no censoring or non-compliance
(we actually observe $Y_{ij} = Y_{ij}(Z_{ij})$ for all $(i, j)$).

Furthermore, we observe a covariate $X_{ij}$ that may be associated with the outcome.
For expository clarity, we suppose that $X$ is univariate, but all results are easily extended to the case when $X$ is multivariate.
$X$ may be associated with stratum membership.

We are interested in the effect of treatment, measured as differences in potential outcomes $Y_{ij}(1) - Y_{ij}(0)$.
We can never learn this difference for any particular individual.
However, a tractable problem is to estimate the mean difference in the study sample or in a target population.
Various functions of potential outcomes may be of clinical interest; the goal of the study and the method of analysis determine which function is considered.

We study hypothesis testing for whether these differences are nonzero using parametric ANCOVA and its permutation counterparts, assuming this potential outcomes framework throughout.
Other valid methods for comparing two groups include using a two-sample $t$ test to test the difference in two means from normal distributions,
the Wilcoxon rank sum test to test for differences in the medians of two independent groups, 
and the Kolmogorov-Smirnov test and receiver operating characteristic curve analyses to test whether two groups have different distribution functions (\cite{lehmann_nonparametrics_1975,  vexler_statistical_2016}).
We focus on testing using the linear model as this is standard in clinical trials, 
requires fewer distributional assumptions on the data when using the potential outcomes framework, 
deals with averages,
and incorporates control variables to increase power.

\subsection{Parametric ANCOVA}\label{subsec:ancova}

ANCOVA is based on a linear model with an indicator variable for membership in each stratum.
The model is

\begin{equation}\label{eqn:ancova}
Y_{ij} = \alpha_j + \beta X_{ij} + \gamma Z_{ij} + \varepsilon_{ij}
\end{equation}

\noindent where $\alpha_j$ is a fixed effect for stratum $j$, $\beta$ is the coefficient for the pretreatment covariate,
$\gamma$ is the coefficient for treatment,
and $\varepsilon_{ij}$ is an error term.
The parameter of interest is $\gamma$, and the parametric ANCOVA tests the null hypothesis $H_0: \gamma = 0$ against
the two-sided alternative hypothesis $H_1: \gamma \neq 0$.
If the linear model is the true data-generating process, then $Y_{ij}(1) = Y_{ij}(0) + \gamma$ for all $(i, j)$.
However, we needn't take this perspective for $\gamma$ to be a useful quantity; it represents the average treatment effect, holding the other variables fixed.

To carry out the standard parametric hypothesis test for a linear model, the following assumptions are needed (\cite{freedman_statistical_2005}):
\begin{itemize}
\item \textbf{Linearity:} The data $Y$ are related to $X$ and $Z$ linearly.
\item \textbf{Constant slopes:} Stratum membership only affects the intercept $\alpha_j$, not the slopes $\beta$ and $\gamma$.
\item \textbf{IID Errors:} The $\varepsilon_{ij}$ are independent and identically distributed with mean $0$ and common variance $\sigma^2$.
\item \textbf{Independence:} If $X$ is random, $\varepsilon$ is statistically independent of $X$.
\item \textbf{Normality:} The errors are normally distributed.
\end{itemize}

The coefficients are estimated using least squares (or equivalently, by maximizing the likelihood).
The coefficient $\hat{\gamma}$ is the estimated average treatment effect. 
This procedure also yields an estimate $\hat{\sigma}_{\hat{\gamma}}^2$ of the variance of $\hat{\gamma}$.
Under the null hypothesis, the test statistic 
$$ T = \frac{\hat{\gamma}}{\sqrt{ \hat{\sigma}_{\hat{\gamma}}^2}}$$
follows the Student $t$ distribution with degrees of freedom equal to the number of observations minus the number of parameters estimated (in this case, $N - J - 2$).
The $p$-value for this hypothesis test is the probability, assuming the null hypothesis of zero coefficient is true, that a value drawn from the $t$ distribution is larger in magnitude than the observed $T$.
This test is equivalent to the $F$ test.
When the model assumptions are true, it is uniformly most powerful.

The linear model is robust to violations of its assumptions, but theoretical guarantees tend to be asymptotic.
When randomized experiments are analyzed in the parametric framework, which assumes that treatment assignment is fixed and the errors are random,
the estimated treatment effect $\hat{\gamma}$ and nominal standard errors $\hat{\sigma}_{\hat{\gamma}}$ can be severely biased (\cite{freedman_regression_2008, lin_agnostic_2013}).
\cite{lin_agnostic_2013} shows that running a regression with a full set of treatment and covariate interaction terms cannot hurt asymptotic precision, and using the Huber-White sandwich standard errors can yield asymptotically valid confidence intervals.
In small samples, the bias may still be substantial.
\cite{miratrix_adjusting_2013} show that post-stratification, estimating treatment effects within groups of similar individuals defined after data are collected, can ameliorate this bias.
This is equivalent to estimating a fully saturated linear model with interaction terms for treatment and stratum membership.
The linear ANCOVA model does not account for variation in the treatment effect across strata in this way;
if the difference in potential outcomes is heterogeneous across strata, then the coefficient $\gamma$ may be attenuated towards 0.
Thus, it is unclear for any particular dataset whether or not the ANCOVA model will give valid results.

\subsection{Stratified permutation test}\label{subsec:strat_perm_test}
All permutation tests essentially have two requirements: a conditioning space, the orbit of a finite group of transformations of the data in which all configurations of the data are equiprobable under the null hypothesis, and a set of sufficient statistics for the data which describe the data under these transformations.(\cite{pesarin_permutation_2010}).
In experiments, the only random quantities are those involving $Z$, the vector of treatment assignments.
In particular, the potential outcomes and covariates are fixed in the finite population under study, while the observed responses change with $Z$.
The exact permutation inference is derived from the conditioning space which includes all possible values of $Z$.

Suppose we wish to test the null hypothesis that individual by individual, treatment has no effect.
This is referred to as the ``sharp'' null hypothesis:

$$H_0: Y_{ij}(1) = Y_{ij}(0), \forall i = 1, \dots, n_j, j = 1,\dots, J.$$

Under the null hypothesis, which treatment an individual received amounts to an arbitrary label.
Once we observe one response under a particular treatment, we can impute the other potential outcome; namely, it would have been the same  (\cite{rosenbaum_covariance_2002}).
Thus, the observed $Y_{ij}$ form a sufficient statistic for the data (\cite{pesarin_permutation_2010}).
This null hypothesis is stronger than the null hypothesis for the parametric ANCOVA, which only asserts that differences in potential outcomes are zero \textit{on average}.

In an experiment with complete randomization, any of the possible allocations of treatment that assign $n_j^T$ out of $n_j$ individuals to treatment 1 in stratum $j$, for each stratum $j=1,\dots,J$, has equal probability.
We can construct the permutation distribution of any statistic under the null hypothesis by imputing the unobserved potential outcomes using the sharp null hypothesis and by re-randomizing treatment assignments using this principle of equal probabilities.
The experimental design specifies the permutation scheme; no additional assumptions are needed to find the distribution of any statistic.
In practice, the $\prod_{j=1}^J {n_j \choose n_j^T}$ possible allocations of treatment may be too numerous to compute the statistic for each one.
Instead, we typically use Monte Carlo simulations to randomly assign treatment and compute the statistic of interest a large number of times to approximate the randomization distribution to a desired degree of accuracy.

To summarize, the assumptions for the stratified permutation test are the following.
\begin{itemize}
\item \textbf{Complete randomization:} the $\prod_{j=1}^J {n_j \choose n_j^T}$ assignments of treatment which assign $n_j^T$ out of $n_j$ individuals to treatment 1 in stratum $j$, for each stratum $j=1,\dots,J$, has equal probability
\item \textbf{No treatment effect under the null:} both potential outcomes are identical and equal to the observed outcome for each individual $(i, j)$
\end{itemize}

The most commonly used statistic is the difference in mean outcomes of subjects who received treatment 1 and the mean outcomes of subjects who received treatment 0.  
This statistic is unbiased over all possible random assignments of treatment which preserve the number of treated individuals,
is interpretable, 
and has convenient theoretical properties owing to it being the difference of two means.
However, the difference in means may not be optimal if we want to detect heterogeneous effects.  
For an extreme example, imagine that the sample contains an equal number of males and females, and each treatment is assigned to half of males and half of females.  
Everybody who receives treatment 0 has an outcome of 0, but males who receive treatment 1 have an outcome of $1$ and females who receive treatment 1 have an outcome of $-1$.  
Then the difference in means between the treatment groups is $0$, even though the treatment had nonzero effects on both males and females.  
This differential effect gets averaged out.

To avoid this, one may want to account for the stratification. 
One way to construct a stratified test is to use the same stratified permutation scheme but a different statistic which combines the stratum-specific statistics into a single value, for instance taking the sum of their absolute values.
Taking the absolute value before summing ensures that effects with different signs do not cancel each other out.
Another way to construct a test is to use the nonparametric combination (NPC) framework proposed by \citet{pesarin_permutation_2010}.
In this framework, the ``global'' null hypothesis of no effect whatsoever is decomposed into the intersection of ``partial'' null hypotheses of no treatment effect within each stratum.
For NPC, one first conducts each partial test separately, then combines their $p$-values in a way that preserves dependencies.
For a randomized experiment, this method is equivalent to combining stratum statistics directly, because treatment is independently assigned across strata.

\subsection{Permutation tests with the linear model}\label{subsec:lm_perm_tests}

The tests in Section~\ref{subsec:strat_perm_test} only use information on the treatment and the outcome.
However, experimenters typically record additional covariates that may be predictive of the outcome.
Incorporating these covariates can reduce variation in the statistic due to factors unrelated to treatment, thereby increasing power to detect a treatment effect.
The permutation tests in this section use the linear model in Equation~\ref{eqn:ancova} to control for covariates, but make different assumptions about the data.
In a randomized experiment, the treatment assignment is independent of covariates, errors, and potential outcomes,
making several variables exchangeable.
We show two permutation tests developed in this framework.
They test the same ``sharp'' null hypothesis of no treatment effect.

The first linear model permutation test is a variation on the stratified permutation test described in Section~\ref{subsec:strat_perm_test}.
The assumptions are the same.
The only difference is the test statistic:
this test uses the $t$ statistic for the regression coefficient of $Z$ in the linear model in Equation~\ref{eqn:ancova}.
Rather than using the $t$ distribution to calculate the significance of the observed $t$ statistic, we construct the permutation distribution of the $t$ statistic by simulation.
Again, the experimental design specifies the randomization distribution and no additional assumptions are needed:
$Z$ is independent of $(Y(1), Y(0), X)$ conditional on strata.

\citet{freedman_nonstochastic_1983} propose an alternative test based on the residuals of the linear regression.
They take an alternative view of the problem:
instead of assuming the data are generated according to the linear model Equation~\ref{eqn:ancova}, they \textit{define} the errors $\varepsilon_{ij}$ to be the difference between the observed outcome $Y_{ij}$ and the data's linear projection onto the plane $\alpha_j + \beta X_{ij}+ \gamma Z_{ij}$.
The $\varepsilon$ are fixed approximation errors in this framework, not independent and identically distributed random errors.

If the null hypothesis is true, then $\gamma = 0$ and $\varepsilon_{ij} = Y_{ij} - \alpha_{ij} - \beta X_{ij}$ for all $i = 1, \dots, n_j, j = 1, \dots, J$.
Therefore, we may estimate the errors by $\hat{\varepsilon} =Y - \hat{Y}$, where $\hat{Y}$ is the vector of fitted values from the regression of $Y$ on $X$ but not $Z$.
The $\hat{\varepsilon}$ approximate the true errors $\varepsilon$ from the true data-generating process, assuming that the null hypothesis is true.
Furthermore, under the null hypothesis, the $\varepsilon$ are independent of $Z$ within strata. 

We can carry out an approximate permutation test by permuting the estimated $\hat{\varepsilon}$ within strata, independently across strata.
The test is only approximate because we must estimate $\alpha$ and $\beta$ to construct $\hat{\varepsilon}$.
The test has two important assumptions pertaining specifically to the linear regression, on top of the two necessary pieces of a permutation test:
\begin{itemize}
\item \textbf{Linear model:} the linear model that describes $Y$ using $X$ and the approximation error $\varepsilon$ holds
\item \textbf{Linear regression fit:} the linear regression residuals $\hat{\varepsilon}$ are a good approximation to the true errors $\varepsilon$ 
\item \textbf{Exchangeable errors:} the $\varepsilon$ are exchangeable against $X$ within strata; any configuration of the $\varepsilon$ amongst the units has the same probability
\item \textbf{No treatment effect under the null:} the $\varepsilon$ do not depend on $Z$
\end{itemize}

The $\hat{\varepsilon}$ may not even be approximately exchangeable within strata if the regression does not describe the relation between $Y$ and $X$ well.
For this reason, \cite{freedman_nonstochastic_1983} advise against using the method if there are large outlier values in the covariate $X$ or when $X$ and $Z$ are highly colinear.
Randomization does not guarantee the exchangeability against $X$, so this assumption should be checked.
One way, suggested by \cite{freedman_nonstochastic_1983}, is to visually inspect a scatterplot of the residuals against $X$.
One may also test for association between $X$ and $\hat{\varepsilon}$, such as a stratified permutation test using the correlation as a statistic, independently within each stratum and use the nonparametric combination of tests to obtain a single $p$-value (\cite{pesarin_permutation_2010}).

To summarize the steps for constructing a permutation distribution:

\begin{enumerate}
\item Regress $Y$ on $X$ and stratum indicators, \textit{but not $Z$} to obtain $\hat{Y}$. Estimate the errors by $\hat{\varepsilon} = Y - \hat{Y}$.
\item Permute the $\hat{\varepsilon}$ within strata to obtain permuted errors $\hat{\varepsilon}^\pi$.
\item Construct permuted responses $Y^\pi = \hat{Y}+ \hat{\varepsilon}^\pi$.
\item Regress $Y^\pi$ on $X$, $Z$, and stratum indicators. The test statistic is the $t$ statistic for the coefficient of $Z$.
\end{enumerate}

Others have developed variants on these approximate regression-based permutation tests. 
There is some disagreement on what constitutes an appropriate permutation scheme.
\cite{manly_randomization_2006} proposed randomizing the outcomes $Y$, treating them as though they were randomly assigned to pairs $(X, Z)$ under the null hypothesis.
Permuting data this way corresponds to the null hypothesis that all coefficients are 0 in the linear model, which may not reflect the true relationship between variables.
\cite{kennedy_randomization_1995} proposed a permutation method similar in spirit to \cite{freedman_nonstochastic_1983}, but which differs procedurally.
Both methods attempt to measure the correlation between treatment and unexplained variation in outcomes, but
instead of regressing pseudo-outcomes $Y^\pi$ on covariates to obtain a permutation distribution, \cite{kennedy_randomization_1995} recommends using the $t$ statistic from regressing $Z$ on the permuted residuals $\hat{\varepsilon}^\pi$. 

Several authors have compared these tests theoretically and empirically (\cite{anderson_empirical_1999, anderson_permutation_2001, kennedy_randomization_1996}).
They find that the Freedman-Lane test of residuals is asymptotically equivalent to the ``oracle'' exact hypothesis test which we could conduct if we knew which permutations of $Y$ given $X$ were equiprobable under the null hypothesis (\cite{anderson_permutation_2001}).
This agrees with empirical results, which show that the Freedman-Lane test performs better than other linear model permutation tests in simulations, though its advantage is small (\cite{anderson_empirical_1999}). 
Therefore, throughout the rest of the paper we focus on the two linear regression permutation tests we described in detail: linear regression with permuted treatment assignments and the Freedman-Lane method of permuting residuals.

\section{Simulations}\label{sec:simulations}

We simulated data from a randomized experiment using several different data-generating processes.
The variables included stratification, two randomly assigned treatments, a baseline measure before treatment, and the same variable measured after treatment.
We tested for a treatment effect and compared the empirical power of the tests over repeated random treatment assignments.
We compared the following tests:
the $t$ test from the parametric ANCOVA,
a stratified permutation test using the difference in mean outcomes as the statistic
 (called ``stratified permutation'' in what follows),
a stratified permutation test based on the $t$ statistic from the linear regression of outcome on control variables (called ``LM permutation'' in what follows),
and the Freedman-Lane permutation test.
The simulation code is included in four supplementary files.
The supplementary files additionally consider a stratified permutation test using the mean change scores, defined as the difference between the outcome and baseline measure;
we omit the results here.

\subsection{Continuous outcomes}

In the first set of simulations, we generated continuous potential outcomes based on a latent variable and disturbance terms, both of which we manipulated.
We used two scenarios.
In the first scenario, the treatment effect was homogeneous across strata.
We drew a latent random variate $v_{ij}$ from the uniform distribution on $[-4, 4]$.
In the second scenario, the treatment effect was heterogeneous across strata, and we drew the latent random variable according to
\begin{displaymath}
   v_{ij} \sim \left\{
     \begin{array}{lr}
       \text{Unif}[-4, -1] & : j=1\\
       \text{Unif}[-1, 1] & : j=2\\
       \text{Unif}[1, 4] & : j=3\\
     \end{array}
   \right.
\end{displaymath} 

We generated independent and identically distributed errors $\varepsilon_{ij}$ and $\delta_{ij}$ and varied the error distribution. 
The errors were either standard normal (to mimic the usual ANCOVA assumptions),
$t$ distributed with two degrees of freedom,
standard lognormal,
or exponentially distributed, with scale parameter $1$ and shifted to have mean zero.
The observed $(v_{ij}, \varepsilon_{ij}, \delta_{ij})$ were independent across $i$ and $j$.

We included $n_j=16$ individuals per stratum and treatment assignment was balanced, i.e. 8 people received each treatment in each stratum.
After sampling $(v_{ij}, \varepsilon_{ij}, \delta_{ij})$, we constructed the baseline value for individual $i,j$ as 
$$ X_{ij} = \frac{-\gamma e^{v_{ij}} + e^{v_{ij}/2}}{2} + \varepsilon_{ij}.$$

\noindent $\gamma$ was a fixed constant.
Then we generated potential outcomes as

$$Y_{ij}(Z_{ij}) = \frac{(2Z_{ij} - 1)\gamma e^{v_{ij}} + e^{v_{ij}/2}}{2} + \delta_{ij}.$$

\noindent This is equivalent to the following linear data-generating process:

$$Y_{ij}(Z_{ij}) = Z_{ij} \gamma e^{v_{ij}} + X_{ij} + (\delta_{ij} - \varepsilon_{ij}).$$

\noindent This model does not exactly match the ANCOVA Equation~\ref{eqn:ancova} and only some of the assumptions are satisfied:
$Y$ is linear in $(X, Z)$ conditionally on the latent variable $v$, 
the errors are independent and identically distributed, and are independent of $X$.
When the latent variables vary in distribution across strata, the assumption of constant slopes is violated.

We regenerated $Z$ and the potential outcomes $Y(Z)$ 10,000 times.
We repeated this procedure for each distribution of latent variables $v$ and of the errors $\varepsilon$ and $\delta$.
The treatment effect for individual $(i, j)$ was $\gamma e^{v_{ij}}$.
We tested whether the treatment had a nonzero effect in the sample using each of the tests described.

\begin{figure}[h]
\centering
\includegraphics[width = \textwidth]{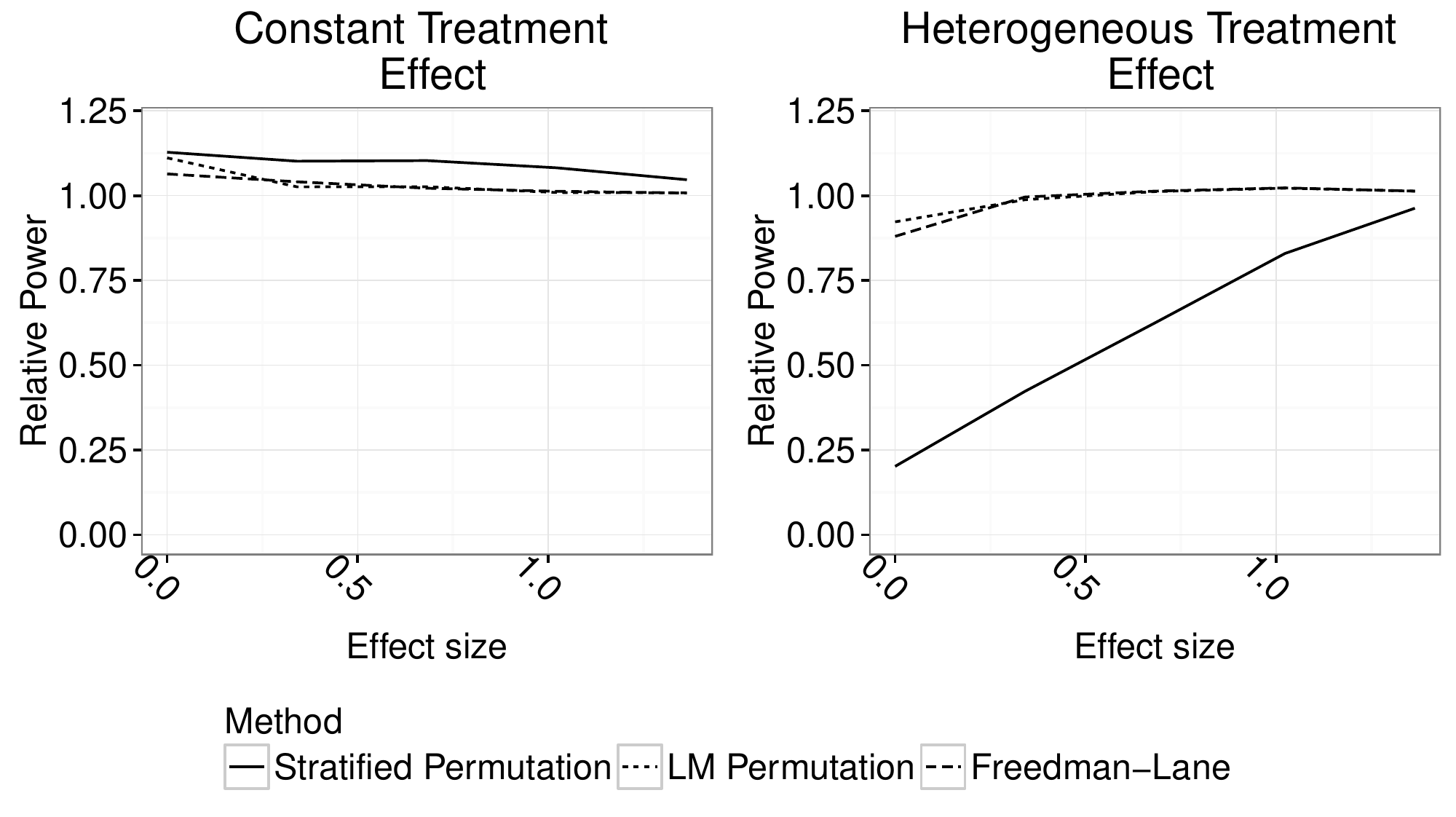}
\caption{Empirical power ratio curves for the continuous simulated data with normally distributed $\varepsilon$ and $\delta$, showing the ratio of the rejection rate of the test to the rejection rate of the parametric ANCOVA. The left panel shows power when the treatment effect was constant across strata and the right panel shows power when the treatment effect was heterogeneous.}
\label{fig:continuous_outcomes_sim_power}
\end{figure}

First, we studied how power changed with the average treatment effect.
We used normally distributed errors and varied $\gamma$ from $0$ to $0.2$ in steps of $0.05$.
In the random populations that were generated, this corresponded to population average treatment effects of $0, 0.34, 0.68, 1.02$, and $1.36$ respectively.
When the treatment effect was homogeneous across strata, all the tests had the correct level of 5\% (i.e. the power when $\gamma = 0$).
The left panel of Figure~\ref{fig:continuous_outcomes_sim_power} shows the empirical power (rate of rejection in the 10,000 simulations) at level 5\%, relative to the empirical power of the ANCOVA, for these increasing effect sizes.
A ratio of one indicates that the test had the same power as the ANCOVA, while a ratio above or below 1 indicates that power was higher or lower, respectively, than the power of the ANCOVA.
The permutation linear model tests performed comparably and are plotted on top of each other, while the stratified permutation test had slightly more power.

The right panel of Figure~\ref{fig:continuous_outcomes_sim_power} shows the empirical power ratios at level 5\% for the same experiment, with heterogeneous treatment effects across strata.
Three of the four tests had the correct level of 5\%, while the stratified permutation test rejected only 1\% of the time.
The stratified permutation had lower power than the linear model tests as the effect size increased.
The correlation between baseline $X$ and outcome $Y$ was low but nontrivial (between $0.04$ and $0.44$), so this result makes some sense:
it is beneficial for power to control for the baseline in the linear model.

Finally, we varied the distributions of the errors $\varepsilon$ and $\delta$ and examined the power of the four tests.
Recall that these are not errors in the linear model framework, but rather disturbances on the potential outcomes and baseline measures for each individual.
These disturbances affect the distributions of $X$ and $Y$, which may impact the power of the parametric ANCOVA and may make it beneficial to control for the baseline.
We fixed $\gamma=0.2$ and let the errors have a standard normal distribution, a normal distribution with variance depending on the magnitude of $X$ (either standard deviation of 1 if $\lvert X \rvert \leq 1$ or standard deviation of 2 otherwise), a $t$ distribution with 2 degrees of freedom, a standard log normal distribution, or an exponential distribution with parameter $1$, shifted to have mean zero.

The top panel in Table~\ref{tab:power_grid1} shows the empirical power for each of these error distributions and tests, where the $v_{ij}$ came from a single distribution across strata.
In all cases, the correlation between baseline and outcome was lower than 0.05 in magnitude.
Thus, the stratified permutation test, which omits the baseline measurement, had the highest power for all error distributions.
All three linear model based tests had roughly the same rejection rate.
Power was highest for the homoskedastic normally distributed errors and was lowest for the heteroskedastic normally distributed errors.
The scenario with correlated errors violates both the ANCOVA assumption of IID errors and the Freedman-Lane test assumption that errors are exchangeable against $X$.

The bottom panel of Table~\ref{tab:power_grid1} shows the results when the distribution of $v_{ij}$ varied across strata.
In this case, the three linear model based tests had the highest power for each error distribution.
While the correlation between baseline and outcome was still nearly zero, controlling for it in the linear model tended to increase precision.
Conversely, the stratified permutation test lost power for this reason.
Interestingly, when the errors were $t$ or log normally distributed, the power of each test did not change much whether the treatment effects were constant or heterogeneous across strata.
Perhaps surprisingly, the power of the linear model tests was higher when effects were heterogeneous than when they had the same distribution across strata.
In this scenario, the linear model does not fully describe the relationship between baseline and outcome: 
interaction terms between baseline and stratum ID are needed in the model to capture this variation across strata.

\begin{table}[ht]
\centering
\begin{adjustbox}{width=\textwidth}
\begin{tabular}{L | l |LLLL}
  \hline
Treatment Effect & Errors & ANCOVA & Stratified Permutation & LM Permutation & Freedman-Lane \\ 
  \hline
\multirow{4}{*}{Constant} & Normal & 0.835 & 0.874 & 0.842 & 0.841 \\ 
 & Normal, heter. & 0.503 & 0.541 & 0.509 & 0.511 \\  
 & t(2) & 0.440 & 0.481 & 0.448 & 0.450 \\ 
 & Log Normal & 0.687 & 0.709 & 0.696 & 0.698 \\ 
 & Exponential & 0.780 & 0.820 & 0.793 & 0.791 \\ 
   \hline
   \hline
\multirow{4}{*}{Heterogeneous} & Normal & 0.877 & 0.844 & 0.887 & 0.891 \\ 
  & Normal, heter. & 0.633 & 0.631 & 0.644 & 0.643 \\ 
  & t(2) & 0.543 & 0.512 & 0.550 & 0.551 \\ 
  & Log Normal & 0.711 & 0.608 & 0.730 & 0.731 \\ 
  & Exponential & 0.814 & 0.786 & 0.820 & 0.816 \\ 
   \hline
\end{tabular}
\end{adjustbox}
\caption{Empirical power at level $0.05$ for simulated data with homogeneous treatment effects (top panel) and heterogeneous treatment effects (bottom panel).} 
\label{tab:power_grid1}
\end{table}

\subsection{Discrete outcomes}

Many medical outcomes are measured on a discrete, ordinal scale, but the same regression-based methods designed for continuous data are used to test for a nonzero treatment effect.
These data do not satisfy the ANCOVA assumptions of linearity and of normally distributed errors.
We repeated the procedure above, instead discretizing the covariate $X$ and potential outcomes $Y(Z)$ by removing their fractional parts.
We used the same distributions for the latent variables $v$.
We drew $\varepsilon$ and $\delta$ from independent standard normal distributions, and again we varied $\gamma$ and the population average treatment effect in the same way.

\begin{figure}[h]
\centering
\includegraphics[width = \textwidth]{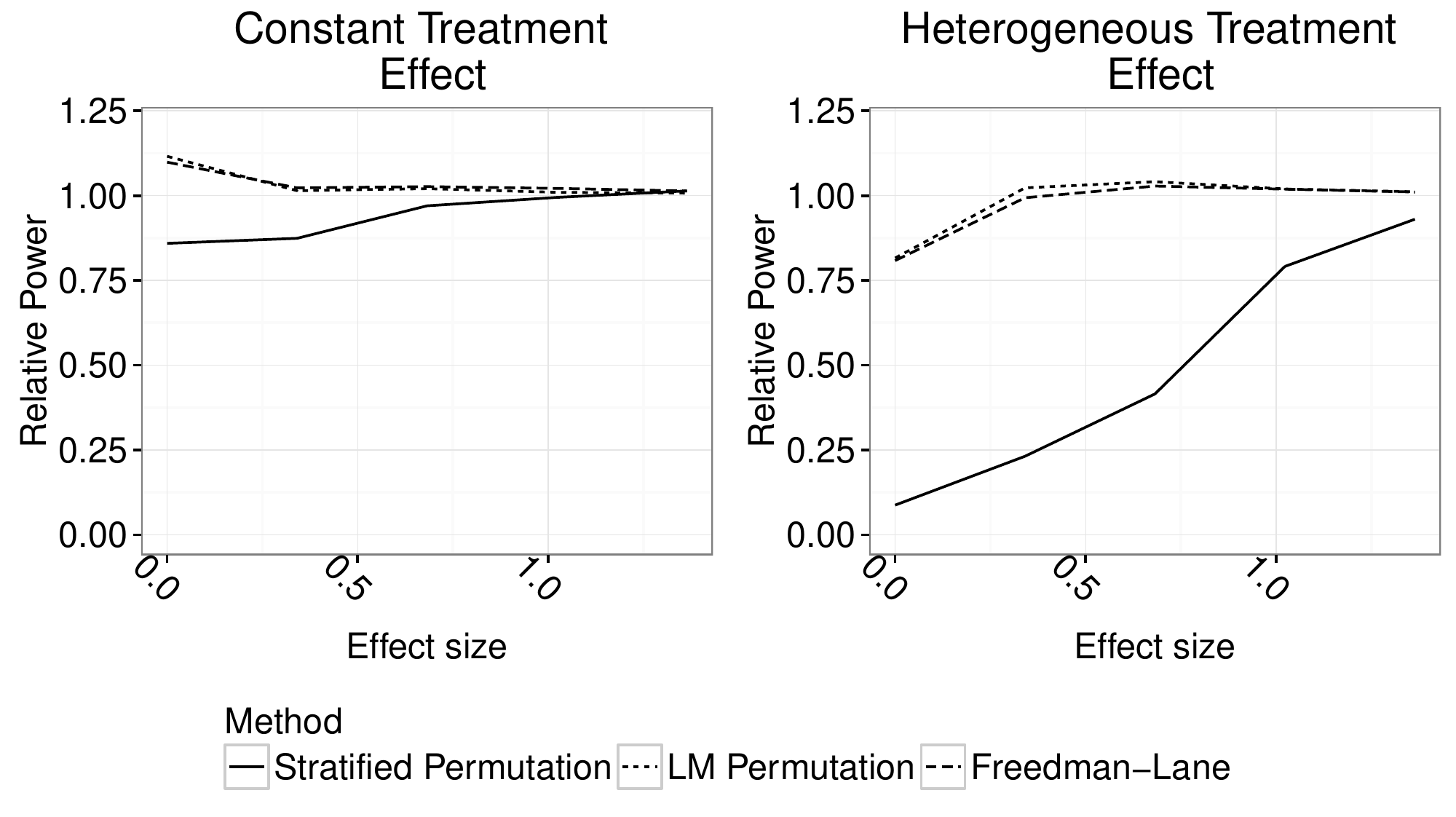}
\caption{Empirical power ratio curves for the discrete simulated data, showing the ratio of the rejection rate of the test to the rejection rate of the parametric ANCOVA. The left panel shows power when the treatment effect was constant across strata and the right panel shows power when the treatment effect was heterogeneous.}
\label{fig:discrete_outcomes_sim_power}
\end{figure}

The left panel of Figure~\ref{fig:discrete_outcomes_sim_power} shows the empirical power at level 5\% relative to the empirical power of the ANCOVA for these increasing effect sizes when treatment effects were homogeneous.
When the effect size was zero, all four tests had the correct level of 5\%, though the stratified permutation test rejected 3.5\% of the time.
In this case, the unadjusted stratified test had comparable power to the linear model based tests.
The right panel of Figure~\ref{fig:discrete_outcomes_sim_power} shows the empirical power ratios when treatment effects varied across strata.
The patterns are similar to when outcomes were continuous.
It is interesting to note that for the discrete outcomes, all of the tests had more power when treatment effects were heterogeneous.

\subsection{Imbalanced treatment groups}
Many experiments use assign a different number of individuals to each treatment group.
Our previous simulations assumed that the number in each treatment group was the same.
We examined the effect of varying the proportion who received treatment in each stratum.
The tests remain consistent no matter how many individuals are assigned to each treatment group, 
but power may be affected.

We repeated the simulations with continuous outcomes, Gaussian disturbances, and fixed $\gamma=0.2$ but varied the number of treated in each stratum to be $4$, $8$, or $12$.
Each of the tests tended to have higher power when the total number of treated units was around half of the population (in this case, 24 out of 48).
It is interesting to note that all tests had more power when 25\% of units were treated than when 75\% were treated; it seems that power is not a symmetric function.
In the scenarios with a varying treatment effect, the  effect was largest in the third stratum and smallest in the first stratum.
As expected, among experiments that fixed the total number of treated units, the experiment with more treatment units in the third stratum tended to have higher power.

The unadjusted stratified test had comparable power to or higher than the linear model based tests.
ANCOVA had slightly higher power than the linear model permutation tests for all experiments except for when the number of treated units was equal to number of controls and when the number of treated units was 30 or above.
Overall, these differences were not practically significant.
We refer the reader to the supplementary file for the details of these experiments.

\subsection{Nonlinear model}

The potential outcomes model we used before turned out to be a linear function of $X$ and $Z$, conditional on the latent variable $v$.
For these simulations, we changed the outcome model to be nonlinear, violating an important ANCOVA assumption.
Instead of an additive treatment effect, now the effect is multiplicative.
We generated the latent variable and errors in the same way, but let

\begin{align*}
X_{ij} &=  \frac{ e^{v_{ij}} + e^{v_{ij}/2}}{2} +\varepsilon_{ij} \\
Y_{ij}(Z_{ij}) &= (1+\gamma)^{Z_{ij}} X_{ij}  + \delta_{ij} \\
&= \gamma X_{ij}Z_{ij} + X_{ij} + \delta_{ij}.
\end{align*}

The treatment effect for individual $(i, j)$ is thus $\gamma X_{ij}$.
The stratified, unadjusted permutation test failed here:
the potential outcomes depend on the baseline covariate, so units are not exchangeable without controlling for the baseline.
The ANCOVA and permutation linear model tests had identical power.
The patterns were the same when the latent variables $v$ had the same and different distributions across strata;
it made little difference because the treatment effect varied within strata due to the dependence on the baseline covariate.

\section{Clinical data results}\label{sec:results}
We compared the parametric ANCOVA, the stratified permutation test, and the two linear model-based permutation tests using a dataset from a clinical trial comparing the effectiveness of two treatments for gastroesophageal reflux disease (GERD).
A detailed discussion of the data and analysis is provided in a supplementary file.
We summarize the analysis here.
Patients were treated at eight sites in two different countries.
At each site, patients were randomly assigned one of two treatments.
Patients were observed for a week before receiving treatment and for a week after receiving treatment.
On each of the fourteen days of observation, patients responded to a survey about their heartburn, regurgitation, and dyspepsia frequency and severity.
These endpoints were measured on a discrete scale.
There were several additional endpoints calculated from the survey measures: daily heartburn, daily regurgitation, daily ``hrdq'' (a composite score), and daily dyspepsia.
Daily ``hrdq'' was the primary endpoint.
To reduce day-to-day variation, we averaged the measures from each week to obtain two observations per patient, one pre-treatment and one post-treatment.
These averages are not quite discrete ordinal values, but they are not continuous either: they can only take a finite set of values.

We used site as the stratification variable, as this is the level at which treatment was randomized.
The model used for the linear regression-based tests was defined as in Equation~\ref{eqn:ancova}.
This model allowed the intercept $\alpha_j$ to vary across sites and used the pretreatment, baseline measurement as the control variable $X$.
The outcome and baseline had a moderate correlation (for instance, the correlation ranged from $0.56$ for daily ``hrdq'' and $0.70$ for heartburn frequency).
\cite{frison_repeated_1992} recommend using change scores only if the correlation is at least $0.5$.
However, we choose to include the baseline as a covariate in the linear regression, rather than using change scores as the dependent variables, to illustrate the inclusion of covariates and to be consistent with the original analysis of the data.

Table~\ref{tab:clinical_distr} shows the mean and standard deviation of each clinical endpoint for the two treatment groups.
There is a large difference in means for daily heartburn (``daily\_heart''), daily ``hrdq'' (``daily\_hrdq''), heartburn frequency (``heart\_freq''), and regurgitation frequency (``regurg\_freq'').
The difference is less clear for daily regurgitation (``daily\_regurg''), daily dyspepsia (``daily\_dysp''), and dyspepsia frequency (``dysp\_freq'').
The distribution of outcomes for each endpoint is extremely right skewed, which may make it difficult to fit a linear model to the data.
Residual plots indicate heteroskedastic errors, invalidating the assumption of constant error variance needed for parametric inference.
The permutation test assumptions are guaranteed by the randomization that was conducted.
We checked the additional exchangeability conditions for the Freedman-Lane test and found no evidence that they were violated. 
These checks are included in the supplementary file.

Table~\ref{tab:clinical_pvalues} shows the $p$-values for the four tests and the seven continuous study endpoints.
Overall, the results confirm our expectations based on visual comparison in Figure~\ref{fig:clinical_distr}:
one or more of the tests reject the null hypothesis that outcomes are the same between treatments for heartburn frequency, daily heartburn, and daily ``hrdq,''
but not for any of the other endpoints.
The $p$-values for the stratified, unadjusted permutation test have no consistent pattern: sometimes they are smaller than the $p$-values from the other tests and sometimes they are larger.

The three tests based on the linear model give qualitatively similar results here.
The ANCOVA $p$-values tend to be smaller than the permutation linear model and residual permutation test $p$-values.
This may be due to the skewed outcome distributions and heteroskedastic errors: the parametric test is more sensitive to large values than the permutation tests.
Our conclusions for the heartburn frequency, daily heartburn, and daily ``hrdq'' endpoints differ between ANCOVA and the permutation tests at significance level $0.05$, but not at level $0.1$.
There is no endpoint which would be deemed significant using a permutation test but insignificant with ANCOVA.
This suggests that the parametric test correctly discriminates between endpoints that are significantly different and endpoints that are not different between treatment A and treatment B.
The correspondence between parametric and permutation test results gives some confidence that the results are stable and reliable.

\begin{table}\label{tab:clinical_distr}
\centering
\begin{tabular}{r|cc|cc}
  \hline
 & Mean A & SD A & Mean B & SD B \\ 
  \hline
daily\_heart & 0.59 & 0.46 & 0.75 & 0.52 \\ 
  daily\_regurg & 0.52 & 0.44 & 0.58 & 0.45 \\ 
  daily\_dysp & 0.56 & 0.51 & 0.55 & 0.49 \\ 
  daily\_hrdq & 1.11 & 0.84 & 1.33 & 0.86 \\ 
  heart\_freq & 2.48 & 2.10 & 3.51 & 2.76 \\ 
  regurg\_freq & 2.30 & 1.97 & 2.58 & 2.16 \\ 
  dysp\_freq & 2.49 & 2.48 & 2.36 & 2.25 \\ 
   \hline
\end{tabular}
\caption{Mean of each continuous outcome in groups A and B.} 
\end{table}

\begin{table}[]
\centering
\label{tab:clinical_pvalues}
\begin{tabular}{l|LLLL}
Endpoint      & ANCOVA & Stratified Permutation & LM Permutation & Freedman-Lane \\ \hline
heart\_freq   & 0.035                      & 0.003                                                                                & 0.070                                                                        & 0.074                                                                       \\
regurg\_freq  & 0.136                      & 0.157                                                                                & 0.231                                                                        & 0.221                                                                        \\
dysp\_freq    & 0.565                      & 0.925                                                                               & 0.592                                                                        & 0.579                                                                        \\
daily\_heart & 0.032                      & 0.010                                                                                & 0.062                                                                        & 0.069                                                                        \\
daily\_regur  & 0.142                      & 0.195                                                                                & 0.250                                                                        & 0.243                                                                        \\
daily\_hrdq   & 0.043                      & 0.033                                                                                & 0.086                                                                        & 0.088                                                                        \\
daily\_dysp   & 0.582                      & 0.803                                                                                & 0.686                                                                        & 0.699                                                                        \\
\hline
\end{tabular}
\caption{Comparison of p-values from four tests, for each continuous endpoint.} 
\end{table}

\section{Discussion}\label{sec:discussion}

This paper adds to the literature comparing parametric and nonparametric tests.
We simulated a variety of data-generating processes under the potential outcomes model, ranging from the ideal case when treatment effects are homogeneous for all individuals, to cases when errors are heavy-tailed or heteroskedastic, data are discrete, effects vary across strata, and treatment assignment is imbalanced within groups.
The linear regression based tests had comparable power in all circumstances.
They all suffered a loss of power when the linear model was a poor fit to the data.
The stratified permutation test appears to perform best when the correlation between the outcome and covariates is low, while using change scores as the dependent variable in the stratified permutation test works best when the correlation is high.
The linear model permutation tests are a middle ground between these two extremes: they allow one to incorporate covariates with any amount of correlation, assuming that the linear model has a reasonable in-sample fit.
It is a matter of taste which test one chooses for their experiment: while the parametric test may be robust to violations of its assumptions, it seems reassuring that the permutation test can exactly match the randomization that was done without additional assumptions.
Applying a linear model based permutation test as a secondary analysis can give insight into how strongly violations of the ANCOVA assumptions affect results
and give confidence that inferences based on ANCOVA are reliable.

On the one hand, some argue that violations of parametric test assumptions necessitate the use of permutation methods.
\cite{ludbrook_why_1998} point out that medical trials rarely follow the population sampling model that is implicit in parametric methods, while permutation tests condition on the sample at hand.
Many people recommend using permutation tests in place of common parametric tests, such as ANOVA and generalized linear models (\cite{still_approximate_1981, winkler_permutation_2014}).
They argue that there are a myriad of ways that the data may violate the necessary assumptions for the test, and so permutation tests are more robust.

However, parametric and nonparametric tests seem to perform similarly when compared side-by-side in simulations, even when data violate the assumptions of the parametric method.
Medical trials often use Likert scales to score pain or symptom severity, resulting in discrete data that does not match the normality assumptions of parametric tests.
However, \cite{winter_five-point_2010} found that the two sample $t$ test and Mann-Whitney test had comparable Type I and II error rates for five-point Likert scale data, suggesting that the violation of normality does not entirely invalidate the parametric test.
\cite{vickers_parametric_2005} compared the parametric ANCOVA to the Mann-Whitney rank test in the context of randomized experiments, finding that except in extreme situations, ANCOVA was more powerful than the nonparametric test.
Most similar to our question of study, \cite{anderson_empirical_1999} found little difference between several permutation tests for coefficients in a linear model alongside the parametric $t$ test.
In these situations, the permutation test strengthens conclusions by giving evidence that the parametric test is robust to departures from its assumptions.
Our results match those of \citet{vickers_parametric_2005} and \citet{anderson_empirical_1999}, where parametric ANCOVA and regression-based $t$ tests performed the same or better than the comparable nonparametric tests.

In randomized experiments, permutation tests are always valid as long as they match the randomization scheme that was actually conducted.
The exchangeability of treatments is guaranteed by construction.
In observational studies, one must prove that subjects are exchangeable.
\cite{romano_behavior_1990} warns against using permutation tests naively if items are not truly exchangeable. 
For instance, he points to the case where observations have unequal variances.  
This is a problem in practice as one cannot observe errors; it is a leap of faith to assert that they are homogeneous and therefore exchangeable.
\cite{boik_fisherpitman_1987} illustrates this phenomenon using the traditional $F$ test and its permutation counterpart, 
and demonstrates by simulation that the latter has larger than nominal level when the error variances are unequal.
Randomized experiments mitigate this problem: by definition, treatment assignment is statistically independent of all other variables (possibly conditioning on strata) and in expectation over repeated randomizations, the two groups have equal variances.

It is important to note that the permutation tests described here are exact only in the context of randomized experiments.
Treatment is assigned at random and is therefore statistically independent of the covariates $X$ and the errors $\varepsilon$.
In observational studies, treatment may be associated with $X$, $\varepsilon$, or both, often in a way that is difficult or impossible to model.
The statistical independence guaranteed by randomized experiments enables us to construct permutation distributions while holding $X$ fixed.
When exchangeability doesn't hold, we cannot disentangle the effect of $Z$ from the effect of $X$.
One must provide evidence that variables in observational studies are exchangeable in order to achieve an approximately exact test.
The onus is on the researcher to make the case that the variable being permuted is uncorrelated with the other variables being held fixed.
This can be done visually using scatterplots, residual plots, hypothesis tests, and from knowledge of how the data arose and were collected.

Our simulations demonstrate that the method of controlling for baseline covariates matters.
The naive way to control for repeated measures is to use change scores, the difference between the outcome and baseline measures, as the dependent variable.
Simulations in the supplementary files confirm the suggestion of \cite{frison_repeated_1992} to use change scores only when there is a strong correlation between baseline and outcome.
Weak correlations between baseline and outcome occur often in practice, as was the case with our GERD dataset.
When the correlation is weak, the test of change scores may be less powerful than ignoring the baseline altogether.
Instead, we suggest incorporating the baseline in a regression model.
This is more general than using differences;
treating the change scores as the dependent variable is a special case of the linear regression that constrains the coefficient of the baseline measure to be $1$.
The regression approach is more flexible and demonstrably more powerful.

\bigskip
\begin{center}
{\large\bf SUPPLEMENTARY MATERIAL}
\end{center}
All files are also available at \url{https://github.com/kellieotto/ancova-permutations}.

\begin{description}

\item[Continuous outcomes simulations:] Simulations using continuous outcomes in Section~\ref{sec:simulations}, including code and results. (PDF)
\item[Discrete outcomes simulations:] Simulations using discrete outcomes in Section~\ref{sec:simulations}, including code and results. (PDF)
\item[Imbalanced treatment simulations:] Simulations using continuous outcomes with varying proportions receiving treatment in Section~\ref{sec:simulations}, including code and results. (PDF)
\item[Nonlinear model simulations:] Simulations using a nonlinear model of potential outcomes in Section~\ref{sec:simulations}, including code and results. (PDF)
\item[Clinical trial data:] Dataset used in Section~\ref{sec:results} to compare methods. (csv)
\item[README:] Data descriptor file. (txt)
\item[Results:] Detailed explanation, code, and results of comparing methods using the clinical trial dataset in Section~\ref{sec:results}. (PDF)

\end{description}

\bibliographystyle{chicago}

\bibliography{references}
\end{document}